\documentclass[conference,compsoc]{IEEEtran}

\usepackage{svg}
\usepackage{subcaption}
\usepackage{graphicx}
\usepackage[acronym]{glossaries}
\usepackage{booktabs}
\usepackage{xcolor}
\usepackage{multirow}
\usepackage{soul}
\usepackage{layouts}
\usepackage[noend]{algorithm, algpseudocode}
\newcommand{\BfPara}[1]{{\noindent\textbf{#1.}}}

\newcommand{\ignore}[1]{}

\makeatletter
\newenvironment{beamToPath}[1][htb]{%
    \renewcommand{\ALG@name}{BeamToPath}
   \begin{algorithm}[#1]%
  }{\end{algorithm}}
\makeatother

\glsdisablehyper  
\newacronym{AL}{AL}{Aggregation Level}
\newacronym{BPSK}{BPSK}{Binary Phase-Shift Keying}
\newacronym{SDR}{SDR}{Software Defined Radio}
\newacronym{BWP}{BWP}{Bandwidth Part}
\newacronym{CCE}{CCE}{Control Channel Element}
\newacronym{CFI}{CFI}{Control Format Indicator}
\newacronym{CFO}{CFO}{Carrier Frequency Offset}
\newacronym{CORESET}{CORESET}{Control Resource Set}
\newacronym{CP}{CP}{Cyclic Prefix}
\newacronym{CP-OFDM}{CP-OFDM}{Cyclic Prefix Orthogonal Frequency-Division Multiplexing}
\newacronym{CRC}{CRC}{Cyclic Redundancy Check}
\newacronym{C-RNTI}{C-RNTI}{Cell-RNTI}
\newacronym{C-RS}{C-RS}{Cell-specific Reference Signal}
\newacronym{DCI}{DCI}{Downlink Control Information}
\newacronym{UCI}{UCI}{Uplink Control Information}
\newacronym{DMRS}{DMRS}{Demodulation Reference Signal}
\newacronym{gNB}{gNB}{gNodeB}
\newacronym{GUTI}{GUTI}{Globally Unique Temporary Identifier}
\newacronym{IMEI}{IMEI}{International Mobile Equipment Identity}
\newacronym{IMSI}{IMSI}{International Mobile Subscriber Identity}
\newacronym{ISI}{ISI}{Inter-Symbol Interference}
\newacronym{LTE}{LTE}{Long-Term Evolution}
\newacronym{NR}{NR}{New Radio}
\newacronym{NSA}{NSA}{Non-Standalone}
\newacronym{MCS}{MCS}{Modulation and Coding Scheme}
\newacronym{MIB}{MIB}{Master Information Block}
\newacronym{MitM}{MitM}{Man-in-the-Middle}
\newacronym{OFDM}{OFDM}{Orthogonal Frequency-Division Multiplexing}
\newacronym{OSINT}{OSINT}{Open-Source Intelligence}
\newacronym{PBCH}{PBCH}{Physical Broadcast Channel}
\newacronym{PCFICH}{PCFICH}{Physical Control Format Indicator Channel}
\newacronym{PCI}{PCI}{Physical Cell ID}
\newacronym{PDCCH}{PDCCH}{Physical Downlink Control Channel}
\newacronym{PDSCH}{PDSCH}{Physical Downlink Shared Channel}
\newacronym{PUCCH}{PUCCH}{Physical Uplink Control Channel}
\newacronym{PUSCH}{PUSCH}{Physical Uplink Shared Channel}
\newacronym{PRB}{PRB}{Physical Resource Block}
\newacronym{PSS}{PSS}{Primary Synchronization Signal}
\newacronym{QPSK}{QPSK}{Quadrature Phase-Shift Keying}
\newacronym{RAN}{RAN}{Radio Access Network}
\newacronym{RAT}{RAT}{Radio Access Technology}
\newacronym{RB}{RB}{Resource Block}
\newacronym{RE}{RE}{Resource Element}
\newacronym{REG}{REG}{Resource Element Group}
\newacronym{RF}{RF}{Radio Frequency}
\newacronym{RNTI}{RNTI}{Radio Network Temporary Identifier}
\newacronym{RRC}{RRC}{Radio Resource Control}
\newacronym{SA}{SA}{Standalone}
\newacronym{SCS}{SCS}{Subcarrier Spacing}
\newacronym{SFN}{SFN}{System Frame Number}
\newacronym{SNR}{SNR}{Signal-to-Noise Ratio}
\newacronym{SIB}{SIB}{System Information Block}
\newacronym{SIB1}{SIB1}{System Information Block 1}
\newacronym{SIB2}{SIB2}{System Information Block 2}
\newacronym{SIB3}{SIB3}{System Information Block 3}
\newacronym{SIM}{SIM}{Subscriber Identity Module}
\newacronym{SSB}{SSB}{Synchronization Signal Block}
\newacronym{SSS}{SSS}{Secondary Synchronization Signal}
\newacronym{SUPI}{SUPI}{Subscription Permanent Identifier}
\newacronym{SIMD}{SIMD}{Single Instruction Multiple Data}
\newacronym{SUCI}{SUCI}{Subscription Concealed Identifier}
\newacronym{TMSI}{TMSI}{Temporary Mobile Subscriber Identity}
\newacronym{TO}{TO}{Timing Offset}
\newacronym{UE}{UE}{User Equipment}
\newacronym{URLLC}{URLLC}{Ultra-Reliable and Low-Latency Communications}
\newacronym{DoS}{DoS}{Denial of Service}
\newacronym{NAS}{NAS}{Non-Access Stratum}
\newacronym{CA}{CA}{Carrier Aggregation}
\newacronym{MAC}{MAC}{Medium Access Control}
\newacronym{CE}{CE}{Control Element}
\newacronym{RA}{RA}{Random Access}
\newacronym{RAR}{RAR}{Random Access Response}
\newacronym{BS}{BS}{Base Station}
\newacronym{SDU}{SDU}{Service Data Unit}
\newacronym{PDU}{PDU}{Protocol Data Unit}
\newacronym{RLC}{RLC}{Radio Link Control}
\newacronym{PDCP}{PDCP}{Packet Data Convergence Protocol}
\newacronym{SR}{SR}{Scheduling Request}
\newacronym{SCell}{SCell}{Secondary Cell}
\newacronym{CSI}{CSI}{Channel State Information}
\newacronym{CSI-RS}{CSI-RS}{Channel State Information Reference Signal}
\newacronym{PO}{PO}{PDCCH Order}
\newacronym{PRACH}{PRACH}{Physical Random Access Channel}
\newacronym{SRS}{SRS}{Sounding Reference Signal}
\newacronym{SP}{SP}{Semi-Persistent}
\newacronym{UL}{UL}{Uplink}
\newacronym{DL}{DL}{Downlink}
\newacronym{AN}{AN}{Access Network}
\newacronym{CN}{CN}{Core Network}
\newacronym{IP}{IP}{Internet Protocol}
\newacronym{LCID}{LCID}{Logical Channel ID}
\newacronym{CFRA}{CFRA}{Contention Free RA}
\newacronym{CBRA}{CBRA}{Contention Based RA}
\newacronym{TA}{TA}{Timing Advance}
\newacronym{BSR}{BSR}{Buffer Status Report}
\newacronym{TCI}{TCI}{Transmission Configuration Indication}
\newacronym{OTA}{OTA}{Over The Air}
\newacronym{RN}{RN}{Radio Network}
\newacronym{BFR}{BFR}{Beam Failure Recovery}
\newacronym{COTS}{COTS}{Commercial off-the-Shelf}
\newacronym{RTT}{RTT}{Round Trip Time}
\newacronym{MNO}{MNO}{Mobile Network Operator}
\newacronym{ECDF}{ECDF}{Empirical Cumulative Distribution Function}
\newacronym{TP}{TP}{throughput}
\newacronym{HARQ}{HARQ}{Hybrid Automatic Repeat reQuest}
\newacronym{TPC}{TPC}{Transmission Power Control}

\AtBeginDocument{%
  \providecommand\BibTeX{{%
    \normalfont B\kern-0.5em{\scshape i\kern-0.25em b}\kern-0.8em\TeX}}}

\def\BibTeX{{\rm B\kern-.05em{\sc i\kern-.025em b}\kern-.08em
    T\kern-.1667em\lower.7ex\hbox{E}\kern-.125emX}}

\begin{document}

\title{Unprotected 4G/5G Control Procedures at Low Layers Considered Dangerous
}
\author{\IEEEauthorblockN{Norbert Ludant}
\IEEEauthorblockA{\textit{Northeastern University} \\
Boston, USA \\
ludant.n@northeastern.edu}
\and
\IEEEauthorblockN{Marinos Vomvas}
\IEEEauthorblockA{\textit{Northeastern University} \\
Boston, USA \\
vomvas.m@northeastern.edu}
\and
\IEEEauthorblockN{Guevara Noubir}
\IEEEauthorblockA{\textit{Northeastern University} \\
Boston, USA \\
g.noubir@northeastern.edu}
}

\maketitle

\begin{abstract}
Over the years, several security vulnerabilities in the 3GPP cellular systems have been demonstrated in the literature.
Most studies focus on higher layers of the cellular radio stack, such as the RRC and NAS, which are cryptographically protected. 
However, lower layers of the stack, such as PHY and MAC, are not as thoroughly studied, even though they are neither encrypted nor integrity protected. Furthermore, the latest releases of 5G significantly increased the number of low-layer control messages and procedures.
The complexity of the cellular standards and the high degree of cross-layer operations, makes reasoning about security non-trivial, and requires a systematic analysis. We study the control procedures carried by each physical channel, and find that current cellular systems are susceptible to several new passive attacks due to  information leakage, and active attacks by injecting MAC and PHY messages.
For instance, we find that beamforming information leakage enables fingerprinting-based localization and tracking of users.
We identify active attacks that reduce the users' throughput by disabling RF front ends at the UE, disrupt user communications by tricking other connected UEs into acting as jammers, or stealthily disconnect an active user. We evaluate our attacks against COTS UEs in various scenarios and demonstrate their practicality by measuring current operators' configurations across three countries.
Our results show that an attacker can, among other things, localize users with an accuracy of 20 meters 96\% of the time, track users' moving paths with a probability of 90\%, reduce throughput by more than 95\% within 2 seconds (by spoofing a 39 bits DCI), and disconnect users.
\end{abstract}

\maketitle
\thispagestyle{plain}
\pagestyle{plain}
\section{Introduction}
\label{sec:intro}

Due to the increasing pervasiveness of cellular communications, and the addition of critical use cases, security has become a focal point in cellular systems. Researchers continue to show that deployed networks are susceptible to attacks that compromise security, privacy, and availability guarantees. Many of these attacks target the control plane, which manages critical tasks for the operation of the network, such as initial access, authentication, or \gls{UE} state management. The control plane protocol layers that are responsible for these procedures are \gls{RRC} and \gls{NAS}, which are secured by encryption and integrity protection. Despite these protections, several attacks were demonstrated, leading to user tracking, message spoofing, and \gls{DoS}~\cite{touchinguntouchables, LTEphonecatcher,practicalattacks,breakinglayertwo}, in particular leveraging pre-authentication messages~\cite{rootofallevil}. As a result, considerable efforts were devoted to further securing procedures at the \gls{RRC} and \gls{NAS} layers. This materialized in 5G with the encryption of the initial \gls{NAS} message, the introduction of the \gls{SUCI}, and the mandate for integrity protection~\cite{ts33501}.

However, the demand for lower latency  caused a push of several control plane mechanisms to the unprotected lower layers of the 4G/5G protocol stack, i.e., the \gls{MAC} and PHY layers, allowing for faster reconfiguration of the network. For instance, \gls{CA}, a technique that boosts throughput by aggregating carriers at different frequency bands, is enabled and disabled through a special \gls{MAC} layer structure, called \gls{MAC} \gls{CE}. Similarly, the physical layer includes control mechanisms, such as Beam Reporting to facilitate beamforming, or PDCCH Order to trigger \gls{UE} re-synchronization. According to the 3GPP standard, the number of these control messages has significantly increased in the last decade. The first LTE release in 2008 contained 7  \gls{MAC} \glspl{CE}, whereas the first 5G release in 2018 contained 19 different \gls{MAC} \glspl{CE}, a number surpassing 50 in the latest 2023 release. This raises a security concern, as the lower layers are not encrypted nor integrity protected, and can be exploited by an adversary, targeting even connected and authenticated users.

In this paper, we study the security and privacy vulnerabilities of unprotected lower layer procedures and control elements. We find new, passive and active, attacks stemming from security design flaws. For instance, in beam management procedures, that enable a passive attacker to localize and track users by exploiting the spatial beamforming configuration. Our analysis  reveals the potential of very low-power active attacks, that spoof PHY and \gls{MAC} layer control information in broadband cellular networks.  For instance, we find that by injecting MAC CEs and PHY control messages, an attacker can: a) disrupt the random access mechanism, bar network access, and disconnect active users, b) force connected UEs to perform wideband UL transmissions (even when they do not have pending traffic) on radio resources selected by the adversary, creating collisions and jamming other legitimate UEs' transmissions, c) disrupt the ACK mechanism creating multiple retransmissions, d) drain the battery of UEs by activating \gls{CA} for long periods of time, e) trigger network flooding, and f) cause various service degradation attacks. For ethical reasons, we evaluate our active attacks in an isolated testbed with COTS UE, and our passive attacks against our own devices using commercial operators  with no impact on other UEs or the network.

Our passive attack results show that it is possible to localize users with an accuracy below 20 meters 96\% of the time by listening to the random access exchange, and track the movement of active users with a success rate of 90\% by listening to channel state reports. These attacks require a single device and no hardware calibration, as they rely only on information reported by the target \gls{UE}.
Results in our testbed show that our active attacks can effectively block initial access to a cell and disconnect users, barring them from re-connecting and impact other active users' throughput by 79\%. Moreover, by spoofing uplink grants to connected users, we find that we can create an amplified distributed DoS attack where legitimate UEs are tricked to constantly transmit on uplink resources selected by the adversary, even when they have no pending data (due to required padding). This results in an effective jamming of legitimate UEs, disrupting communication and dropping their performance by more than 95\% within 2 seconds. Finally, we find that spoofing MAC CEs, such as Carrier Aggregation activation, can drain users' battery and halve users' throughput. To validate the impact on real networks, we experimentally analyze the current configuration of six major \glspl{MNO} across three countries with over 500 million subscribers, and find that their parameter choices enable these amplification attacks.

\section{Background}
\label{sec:background}
In this section, we summarize the key aspects of the 4G and  5G radio protocols necessary to describe the vulnerabilities we have identified in our analysis.

\subsection{Physical layer}
\label{sec:phylayer}
5G NR was developed as an evolution of the 4G LTE \gls{RAN}. As such, in 5G the same radio-design principles are employed, with an emphasis on flexibility, to adapt to new types of available \glspl{UE} and services. Both technologies use OFDM, and the time and frequency resources are divided in a grid,
with frames, subframes, slots, and OFDM symbols in the time domain, and subcarriers in the frequency domain.
The grid accommodates multiple physical channels, for different purposes. Figure~\ref{fig:gridandstack} shows the spectrogram of a 5G \gls{DL} transmission recording, labeled with some of the important DL physical channels. For instance, the \gls{SSB} provides a means for a UE to obtain time and frequency alignment with the \gls{BS}. 

User and control plane data for a \gls{UE} in the \gls{DL} are carried over the \gls{PDSCH}. The \gls{PDSCH} is a wide, shared spectrum region and each \gls{UE} is notified by the \gls{BS} of \gls{DL} resources addressed to it. This scheduling information is transmitted over the \gls{PDCCH}, in \gls{DCI} messages, and includes time and frequency allocations, modulation and coding scheme, etc.
For \gls{UL} transmissions the procedure is analogous: the \gls{UE}
is notified of \gls{UL} grants via \gls{DCI} messages, i.e., where in the \gls{PUSCH} it
should transmit its data. \glspl{UE} in the \gls{RAN} are uniquely identified and addressed by their \gls{RNTI}.
The \gls{RNTI} is assigned during initial access upon completion of the \gls{RA} procedure,
and is updated if the \gls{RRC} connection is re-established, for instance, after an inactivity period.
\gls{RA} may also be triggered during the \gls{UE} ``Connected'' state to synchronize with the \gls{BS} or request \gls{UL} grants.

\begin{figure}
\centering
\includegraphics[width=0.35\textwidth]{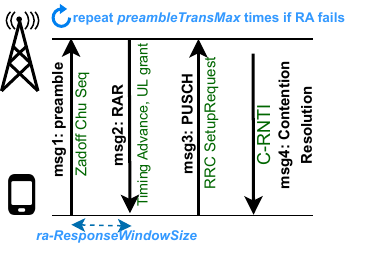}
\caption{Random Access (RA) procedure including the most important information carried by each message.}
\label{fig:RAprocedure}
\end{figure}

The RA procedure is depicted in Figure~\ref{fig:RAprocedure}. To initiate the \gls{RA}, the \gls{UE} chooses randomly from a set of preambles and
transmits it over the \gls{UL} random access channel (\gls{RA} \texttt{Msg1}). Upon receipt of \texttt{Msg1},
the \gls{BS} responds in the \gls{DL} with \texttt{Msg2}, a \gls{RAR}.
The \gls{RAR} includes important information, namely a \gls{TA} Command, which is used to synchronize the UE uplink transmissions, and an \gls{UL} grant
for the \gls{UE} to transmit the \gls{RA} \texttt{Msg3}, whose content depends on the purpose of the \gls{RA}. The limited amount of \gls{RA} preambles in \texttt{Msg1} may lead to collisions between
different \glspl{UE} initiating \gls{RA} at the same time. This is resolved by the \gls{BS} by attaching a ``Contention Resolution''
in \gls{RRC} Connection Setup Complete (\texttt{Msg4}) which completes the \gls{RA}.
The \gls{RA} parameters are included in the \gls{SIB}, which is broadcast by the \gls{BS} few times per second. These
parameters include the number of retries after \gls{RA} failure (\texttt{preambleTransMax}), the power ramp-up after each failure
(\texttt{powerRampingStep}), the \gls{RAR} reception window (\texttt{ra-ResponseWindowSize}), and more.

Finally, as beamforming technologies matured, additional adaptations to the physical layer were introduced in 5G to support its operation. The use of narrow, directional beams uses new procedures between the base station and the UEs for beam management. For instance, beam measurement, beam reporting and beam tracking are included in the physical layer due to their low latency requirements.

\subsection{Cellular 4G/5G protocol stack} 
\label{sec:protocolStack}
\begin{figure}
\includegraphics[width=0.45\textwidth]{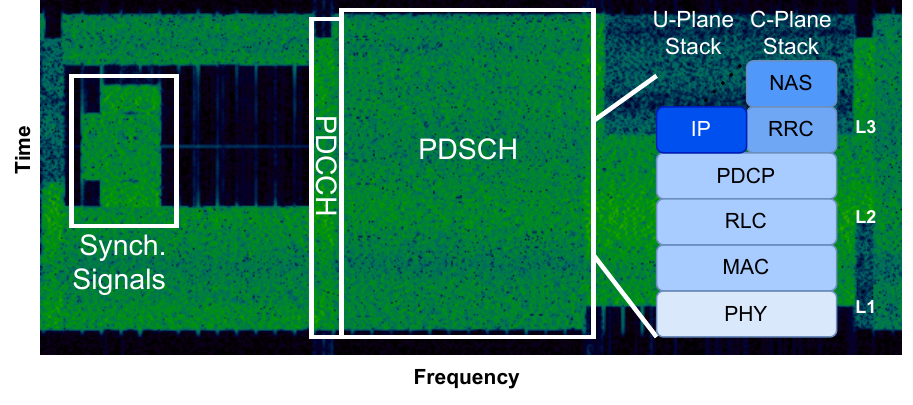}
\caption{Recording of a 5G downlink transmission, with the main physical channels highlighted, and the protocol stack carried by the PDSCH.}
\label{fig:gridandstack}
\end{figure}

The control and user plane protocol stacks carried by the PDSCH are depicted in Figure~\ref{fig:gridandstack}. The access network control plane protocol stack is divided in three layers: (a) L1 which contains PHY, (b) L2 which contains \gls{MAC},
\gls{RLC} and \gls{PDCP}, and (c) L3 which contains RRC. \gls{NAS} is the control protocol between UE and \gls{CN}, and
is not considered part of the radio access network. The L1 and L2 layers are common for the user and control plane stacks, and user data is carried on top of \gls{PDCP}. We briefly describe each layer, with special emphasis on the lower-layer components of L2.

\subsubsection{Radio Resource Control (RRC)}
\gls{RRC} manages the radio connection between \gls{UE} and \gls{BS}.
It is responsible for tasks such as mobility management, user radio configuration, and key management.
This layer is encrypted and integrity protected after successful mutual authentication. As part of the RRC configuration, \gls{BS} and \gls{UE} negotiate which
technologies and mechanisms are enabled.

\subsubsection{Packet Data Convergence Protocol (PDCP) and Radio Link Control (RLC)}
\gls{PDCP} is responsible for IP header compression, encryption, integrity protection, and transport of higher layer user and control data.
The \gls{RLC} layer is in charge of fragmentation or
concatenation of upper layer data to be transported by the \gls{MAC}.

\subsubsection{Medium Access Control (MAC)} The \gls{MAC} layer is the link between the physical and higher protocol
layers, and is in charge of resource scheduling and mapping between logical and transport channels. A single transport channel carries multiple logical channels, for instance, the \gls{DL} shared channel can contain traffic or control data.
To facilitate this, the MAC layer includes a header that indicates which type of logical channel is associated with its
payload, using a field termed \gls{LCID}.

Further, the \gls{MAC} layer carries data and control plane signalling in the MAC header via
\gls{MAC} \glspl{CE}. The purpose of \gls{MAC} \glspl{CE} is to dynamically modify the radio configuration,
complementing the static radio configuration managed by the \gls{RRC}.
The \gls{LCID} header field (index) is therefore overloaded to also convey the type of \gls{MAC} \gls{CE}.
A single \gls{MAC} PDU may carry any combination of zero or more \gls{MAC} \glspl{CE}, \gls{MAC} SDUs, and optionally padding~\cite{ts36321, ts38321}.
A list of MAC CEs can be found in the standard~\cite{ts36321, ts38321}. We describe the security implications of relevant \gls{MAC} \glspl{CE} in Section~\ref{sec:attackoverview}.

\section{Low-Layer attacks}
\label{sec:attack}
We present the adversary model, the L1 and L2 vulnerabilities and attacks, then discuss the practical feasibility of performing said attacks. 
Our analysis encompasses both 4G and 5G technologies. Additional control procedures have been incorporated in the 5G standard to address new functionalities, making certain attacks 5G-specific. We will explicitly state when an attack is only applicable to one of the technologies.

We specifically analyze the information carried by each physical channel: \gls{PBCH}, Physical Downlink and Uplink Control Channels (PDCCH, PUCCH), Physical Random Access Channel (PRACH), and Physical Downlink and Uplink Shared Channels (PDSCH, PUSCH). We classify our findings, based on the type of control information carried by the channel. L2 layer vulnerabilties are described in Section~\ref{sec:L2vulnerabilities}, and the L1 vulnerabilities are described in Section~\ref{sec:L1vulnerabilities}.

Our attacks can be used to passively localize and track user movements, or actively modify the state of a \gls{UE}
by injecting signalling commands at unprotected layers.
This allows an attacker to enable or disable radio techniques at the \gls{UE}, perform amplified distributed jamming attacks, or trigger \gls{UL} wideband radio transmissions and \gls{RA} procedures. This leads to battery draining, performance degradation,
and network access disruption, with low power overhead for the attacker. A summary of our findings is presented in Table~\ref{table:summarizedattacks}, which includes a set of the most important attacks and their impact.

\subsection{Adversarial model}
\label{sec:advmodel}
We assume an attacker positioned within the coverage of the 4G/5G \gls{BS}. The attacker, equipped with a \gls{SDR}, is able to record and process RF I\&Q samples from surrounding cells. We assume that the attacker is also capable of decoding the plaintext \gls{BS} transmissions, as shown previously in ~\cite{OWL,FALCON,lteeye,5gsniffer}. This includes, synchronization signals and system information, scheduling information, and low layer headers, among others. We assume that the attacker can generate and inject 4G/5G wireless signals over the air at desired time instants, as demonstrated by previous research~\cite{adaptOver,LTrack,hidingplainsignal}.

\subsection{Generic attack overview} 
\label{sec:attackoverview}
The attacker listens to the broadcast synchronization signals to achieve time and frequency synchronization with the \gls{BS}.  The attacker then decodes additional unprotected information from the physical channels, such as the \gls{PDCCH}.
This way the attacker deduces the number of connected users, their \glspl{RNTI} and the current load of the \gls{BS}.
The attacker has also access to scheduling information pointing to user-specific data in the resource grid and can eavesdrop unprotected L1/L2 data~\cite{yongdaesniffer}.

As the RF is an open medium and L1 and L2 layers are unprotected, the attacker can either perform passive attacks by listening to the exchange of information between UEs and BS, or active attacks, by spoofing 
messages to specific users, addressed by their \gls{RNTI}.
L1 messages do not contain headers of the protocol stack, and can be spoofed very efficiently by an attacker
by injecting a signal on the physical channel, such as the \gls{PDCCH}.

In order to perform an L2 injection in the \gls{DL}, additional steps are required. The attacker crafts a packet containing L2 data, e.g.,
a \gls{MAC} header, and encodes it in the \gls{PDSCH}. They also craft a \gls{DCI}, containing
the \gls{RNTI} of the user it is addressed to, encoded in the \gls{PDCCH}. This \gls{DCI} points to the location of the crafted L2 message in the \gls{PDSCH}. The attacker, then, transmits the crafted \gls{PDCCH}
and \gls{PDSCH} aligned with the \gls{BS} resource grid at slightly higher power than the \gls{BS}, ensuring overshadowing
of the legitimate \gls{BS} transmissions~\cite{hidingplainsignal}. Alternatively, the adversary can transmit the \gls{PDCCH} over unoccupied resources. 
\gls{UL} attacks follow the same principle: the attacker monitors for \gls{UL} grants in the \gls{DCI} addressed
to a given \gls{RNTI}, and overshadows the \gls{UE} transmission with the crafted L2 header on the \gls{PUSCH}.
When the target receives the crafted message it has no means to verify its legitimacy, and
applies the indicated change.

\begin{table*}[]
\centering
\begin{tabular}{@{}llll@{}}
\multicolumn{2}{c}{\textit{\textbf{PHY Layer (L1) Attacks}}}                                                                 & \multicolumn{2}{c}{\textit{\textbf{MAC Layer (L2) Attacks}}}                                                                                      \\ \midrule
\textbf{Control Procedure} & \multicolumn{1}{l|}{\textbf{Impact}}                                                            & \textbf{Control Procedure}                                                                               & \textbf{Impact}                                                         \\ \midrule
Implicit beam reporting (RA)           & \multicolumn{1}{l|}{Passive user localization} 
& 

\begin{tabular}[c]{@{}l@{}}SP-SRS Act/Deact.\\ CSI-RS Act/Deact.\end{tabular}   & 
\begin{tabular}[c]{@{}l@{}}Passive user localization and tracking \\ Massive MIMO pilot contamination\end{tabular}

\\ \midrule
\begin{tabular}[c]{@{}l@{}} Downlink Control Information \\ Uplink Control Information \end{tabular} 
& \multicolumn{1}{l|}{\begin{tabular}[c]{@{}l@{}}UL jamming by tricked UEs\\ Resource exhaustion, HARQ failure\end{tabular}}

&

SCell Act/Deact. & TP throttling, battery draining 

\\ \midrule

PDCCH Order                & \multicolumn{1}{l|}{\begin{tabular}[c]{@{}l@{}}Targeted DoS\\ Resource exhaustion\end{tabular}}

& 

\begin{tabular}[c]{@{}l@{}} Timing Advance\\ Recommended BitRate \end{tabular} & \begin{tabular}[c]{@{}l@{}}Various QoS effects: \\ DoS, de-synchronization\end{tabular} 
\\ \midrule

BWP Switching              & \multicolumn{1}{l|}{DoS, low-power MiTM enabler}                                 

&

Beam Failure Recovery      & \multicolumn{1}{l}{DoS}    
                                                       
\\ \bottomrule  
\end{tabular}
\caption{Summary of newly discovered L1-L2 layer attacks found from our analysis of the 3GPP cellular standard. The attacks are organized by protocol layer, and sorted in a descending order of relevance: a combination of attack impact and feasibility i.e., high-impact low-effort attacks are listed on top.}
\label{table:summarizedattacks}
\end{table*}

\subsection{Identified vulnerabilities in L1 procedures}
\label{sec:L1vulnerabilities}

\subsubsection{Physical Broadcast Channel (PBCH)} Base stations broadcast a set of information blocks necessary for initial cell access, namely \gls{MIB} and \gls{SIB}. The former is carried by the \gls{PBCH} in the \gls{SSB}, and it contains common physical layer parameters and indications on how to retrieve further \gls{SIB}s. \gls{SIB}s are transmitted unprotected in the PDSCH and contain information such as random access parameters, energy thresholds to connect to a cell, and more. As these broadcast transmissions are crucial for cell access, researchers have studied attacks on both MIB and SIB. For instance, it has been shown that it is possible to block access to a cell by jamming or spoofing MIB~\cite{sigunder,ltejammingspoofing}, and SIB messages~\cite{hidingplainsignal}. We are interested in exploiting this broadcast information in order to amplify other proposed attacks, such as modifying \gls{RA} parameters configured in SIB, described in Section~\ref{sec:phylayer}.

\subsubsection{Physical Downlink Control Channel (PDCCH)} \label{sec:pdcch}

The \gls{PDCCH} is used to convey resource scheduling information to the \gls{UE}. This is carried in the \gls{DCI}, which contains physical layer resource allocation for the downlink and uplink, along with the required data-encoding information. Additionally, 
\glspl{DCI} contain fields to manage various control procedures of the UE, such as power control, \gls{HARQ} information or trigger \gls{RA}. For clarity, we refer to DCIs with downlink allocations as DL DCI, and DCIs with uplink grants as UL DCI.

\gls{DCI} is the target of numerous passive attacks in the literature, such as localization or traffic fingerprinting, by inferring users' information from the resources scheduled to them~\cite{touchinguntouchables}. We expand on this by analyzing the impact of active attacks on \gls{DCI}, that aim to disrupt communication by spoofing either resource scheduling, or control commands to the \glspl{UE}. \gls{DCI} spoofing is highly energy-efficient for the attacker, as it only occupies a small subset of the subcarriers in the PDCCH region (less than 3 OFDM symbols).

\paragraph{\bf{Attacks on resource scheduling}} \gls{DCI} spoofing against a \gls{UE} can be used to fake the allocation of resources either for the downlink or for the uplink. The former is an important basic block in order to successfully spoof higher-layer (e.g., MAC) data to a \gls{UE} (Section~\ref{sec:L2vulnerabilities}). It offers little value otherwise, because the \gls{UE} would simply fail to decode data in an empty slot or in a slot carrying data for another user.
On the other hand, \gls{UL} resource scheduling proves more interesting, as an attacker can force multiple \glspl{UE} to transmit over the same resources, causing jamming of legitimate users and battery draining.

To perform the attack, the attacker selects one or more connected \glspl{UE}, which we call Induced-Jammer UEs (IJ-UE), and injects an \gls{UL} \gls{DCI} in every time slot, allocating a set of resource blocks to them. This instructs the IJ-UEs to transmit data over the allocated RBs during every time slot. \textit{We discovered that the \glspl{UE} use all the allocated resources regardless of how much pending data they need to transmit.} This derives from constraints related to the design and structure of 3GPP L1 scrambling and interleaving. Therefore, the \glspl{UE} pad their \gls{UL} data to fill all allocated resources. Moreover, to maximize the impact, the attacker uses the \gls{TPC} field in the same \gls{DCI} to instruct the exploited \glspl{UE} to transmit at the maximum power. 

This causes the \gls{SNR} of other connected devices to drop close to or below $0$, and severely impacts their throughput. We evaluate the potential of this attack in Section~\ref{sec:evaluation}.

\paragraph{\bf{\gls{PO}}} \gls{PO} is a special DCI used to instruct a specific connected UE to initiate an \gls{RA} procedure, in order to re-establish synchronization in the \gls{UL} (i.e., update the \gls{TA} value).

This control procedure is initiated by a \gls{DCI} with predefined fields,
indicating it is a \gls{PO} message.

\gls{PO} is the only unprotected control procedure that can trigger \gls{RA} for connected users, as other network-initiated \gls{RA} procedures are triggered through the protected \gls{RRC} layer.
This makes \gls{PO} particularly interesting as it provides an efficient
and stealthy way to instruct a \gls{UE} to perform \gls{RA}. The \gls{RA} procedure involves a 4-message exchange
between \gls{UE} and \gls{BS}, as well as \gls{DL} and \gls{UL} resource allocations. A first exploitation of this vulnerability leads to draining resources by simply injecting a single DCI.
This is due to the limited amount of \gls{RA} resources, causing significant collisions.
Additionally, and more critically for privacy, it can be paired with other vulnerabilities to disconnect users and trigger localization attacks, as we discuss in Section~\ref{sec:prach}.

\paragraph{\bf{\gls{BWP} Switching Attack}} \gls{BWP} is a concept introduced in 5G that enhances efficiency and flexibility
for \glspl{UE} with diverse capabilities (e.g., low-power devices). The total bandwidth is divided in multiple parts within the same band, and the \gls{BS} allocates different parts to different users over time. This
decision is made dynamically, pointing the UEs to the \gls{BWP} where their data will be transmitted. The \gls{BWP} switching procedure is either performed at the \gls{RRC} layer, or through the unencrypted \gls{DCI} at the PHY layer.
In the latter case, the \gls{DCI} contains the \gls{BWP} Indicator Field, which informs the \gls{UE} which \gls{BWP} to monitor.
An adversary can create a DoS attack by spoofing a \gls{DCI} to a user and point the UE to another portion of the band, as indicated by the spoofed BWP Indicator Field. As a UE can only have one active \gls{BWP} at a time, transmissions scheduled in the legitimate BWP will be lost, disrupting the connectivity between UE and \gls{BS}. Moreover, this mechanism can be leveraged for more sophisticated and stealthy attacks, for instance facilitating man in the middle (MITM) attacks, where the user can re-direct the UE across \glspl{BWP} at will, and stealthily make the UE miss specific messages from the \gls{BS}, or inject DL data to the user in an empty BWP.

\subsubsection{Physical Uplink Control Channel (PUCCH)}
\label{sec:pucch}
The PUCCH carries \gls{UCI} which contains any of the following messages: Scheduling Requests, uplink ACKs, or CSI reports. The UE uses the RNTI as part of the UCI encoding process, such that the BS can identify the identity of the sender. We identify the following vulnerabilities in the UCI messages.

\paragraph{\bf{Spoofing Scheduling Request}} This uplink physical layer message is sent from the UE to the BS to request uplink resources. Upon reception of a SR, the base station will allocate resources for a user. Scheduling requests can be spoofed and leveraged by an attacker in three ways: i) maintaining users' RNTI connection active for long periods of time, bypassing the RRC inactivity timer, which enables long-term tracking, ii) requesting resources on behalf of multiple users that do not have pending uplink data, leading to congestion in network resources, and iii) to request an UL DCI for a specific user, and hijack the allocated UL grant to spoof higher-layer data on behalf of the user. The second attack can be used as a stealthier alternative to the resource scheduling attacks presented in Section~\ref{sec:pdcch}. This is because an enhanced BS could monitor the downlink to detect spoofed transmissions but spoofed uplink transmissions are harder to detect, as they might come from legitimate UEs. The third attack is more interesting for an attacker as it can be used on-demand to hijack UL grants and inject MAC layer information for a specific user.

\begin{figure}
\centering
\includegraphics[scale=1]{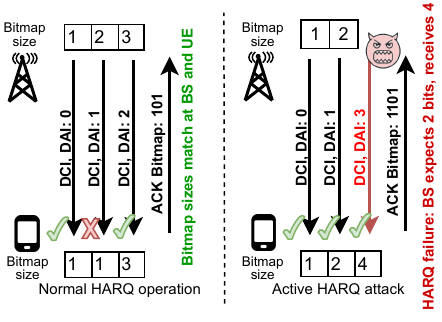}
\caption{Two examples of the HARQ procedure during a missed allocation in normal operation (left) and under an active attack (right). The DAI counter implicitly indicates missed transmissions, and can be leveraged by an attacker.}
\label{fig:harqattack}
\end{figure}

\paragraph{\bf{\gls{HARQ}-Attack}} 
Message acknowledgments (ACKs) are a crucial part of communications. In 4G, the downlink ACKs are transmitted over a predetermined channel, which makes them predictable and susceptible to jamming attacks~\cite{ltejammingspoofing}. 5G omits downlink ACKs altogether, and the UEs are implicitly informed of lost packets by requesting re-transmissions. In the 5G uplink channel, ACKs are scheduled dynamically and can be aggregated in the same UCI. The ACK timing is specified in the DL DCI PDSCH-to-HARQ Feedback Timing Indicator field, that informs the UE of the uplink time slot where the ACK should be reported. The UE reports a dynamic sized bitmap, using one bit for each aggregated ACK. However, if the UE fails to decode a DL DCI, the bitmap size would be mismatched between the BS and the UE. In order to address this scenario, the DL DCI includes a counter (Downlink Assignment Index (DAI)), used by the UE to determine whether any transmissions were lost, and adjust the bitmap size accordingly. This scenario is depicted in Figure~\ref{fig:harqattack} (left). However, these mechanisms do not account for active attackers. An attacker aiming to disrupt the communication of a specific user spoofs a DCI with a modified DAI counter that breaks the synchronization between transmitted and received packets at the UE, as shown in Figure~\ref{fig:harqattack} (right). In this case, when the UE reports the ACK bitmap, it does not match the size expected by the BS, and the BS is unable to resolve to which packets correspond to the ACKs, leading to a \gls{HARQ} failure. We demonstrate this attack in our testbed using a COTS UE in the evaluation section.

\paragraph{\bf{Sniffing \gls{CSI} Reports}} \glspl{UE} perform measurements of the quality of the downlink channel, and report the measured \gls{CSI} to the \gls{BS}. The \gls{BS} uses the CSI parameters as input for resource scheduling. CSI reports are carried by UCI but can be scheduled by MAC layers. We will describe a Layer 2 attack on user tracking  using CSI reports in Section~\ref{sec:srscsi}.

\subsubsection{\gls{PRACH}}\label{sec:prach} The \gls{PRACH} is used for the \gls{RA} procedure, which is the first step to attach to the network. This procedure is unprotected and therefore susceptible to eavesdropping and spoofing. We identify a set of attacks that target the RA procedure.

\paragraph{\bf{Blocking initial cell access}}

The \gls{RA} parameters are broadcast by the BS, to enable new \glspl{UE} to connect to the network. This is facilitated by broadcast in the \glspl{SIB}, as described in Section~\ref{sec:background}. An attacker can overshadow the \glspl{SIB} and modify the \gls{RA} configuration at the \gls{UE} side.
Specifically, the attacker modifies \texttt{ra-ResponseWindowSize} to the minimum value, (i.e., \texttt{sf2}) to shorten the window. This forces the \gls{UE} to only monitor for a \gls{RAR} message within
$3 + \texttt{ra-ResponseWindowSize}$ subframes as described in the 3GPP standard. When this timer elapses, the \gls{UE} deems the \gls{RA} unsuccessful and retries for a maximum of \texttt{preambleTransMax} attempts. To increase congestion, the spoofed \glspl{SIB} also set the \texttt{preambleTransMax} to the maximum value (i.e., 200). In the evaluation section, we show that the configuration of analyzed commercial operators renders them even more susceptible to this attacks. 

This attack causes the \gls{RA} to fail for all \glspl{UE} that are trying to connect to the network, but does not affect already connected \glspl{UE}, which do not monitor the \glspl{SIB}. To target already connected users, the attacker injects an \gls{SIB} paging which instructs all connected \glspl{UE} to monitor the \gls{SIB} for updates~\cite{hidingplainsignal}. The attacker can now inject a PDCCH Order (PO) DCI to a target \gls{UE}, which triggers the \gls{RA} procedure.

This attack is based on the assumption that the \gls{BS}, unaware of the change in \gls{RA} parameters, does not prioritize the \gls{RAR}, leading to \gls{RA} failure due to \gls{RAR} timer expiration. Further, \glspl{PO} to multiple users can be injected at the same time, since the \gls{PDCCH} can fit many \glspl{DCI}, which amplifies the collision effect during \gls{RA}.
The repeated \gls{RA} attempts flood the random access channel, creating overhead at the \gls{BS}, and hindering the access for new users. Additionally, the UE ramps up the power after every failed \gls{RA} causing increased battery consumption.
As a result, by paging active users and overshadowing the \gls{SIB}, the attacker controls the time of the
attack and operates in a power-efficient and stealthy manner, instead of continuously overshadowing the \gls{SIB}. In Section~\ref{sec:results}, we validate our assumption with a real-world passive measurement campaign of \gls{RAR} times of three major cellular \glspl{MNO} and also demonstrate the active attack on our isolated testbed.

\begin{figure}
\centering
\includegraphics[scale=1.5]{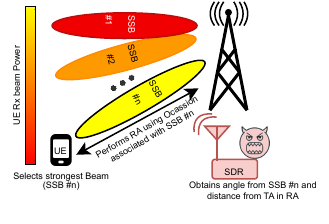}
\caption{The UE measures the strongest beam and implicitly reports it during RA. An attacker passively listening to the exchange, which includes TA, can then localize the user.}
\label{fig:ssbrafig}
\end{figure}

\paragraph{\bf{SSB-RA fingerprinting localization attack}} 5G NR uses beamforming techniques to create directional transmissions (i.e., \textit{beams}) that focus the transmitted signals to specific locations. New mechanisms are introduced in 5G to support beamforming.
Specifically, the \gls{RA} procedure is modified to include an implicit beam-reporting mechanism that facilitates beamforming since the initial message exchange. This procedure is depicted in Figure~\ref{fig:ssbrafig}. The \gls{BS} broadcasts the \glspl{SSB} and \glspl{SIB} over different beams, each identified by a unique index in the cell. The UE measures the received power for each SSB, and determines the strongest beam. 3GPP defines a one-to-one mapping between the SSB beam indexes and RA occasions, such that the BS can determine the optimal beam for the UE based on the RA occasion used by the UE.

This process can be exploited by an attacker, using a single-device fingerprinting-based approach, to localize 5G users. The attacker fingerprints the beam configuration of a cell, and creates a map with the precise locations of the \gls{BS} and every beam in that cell. Then, the attacker monitors the random access channel and deduces the beam chosen by the \gls{UE} from the random access occasion. Additionally, the attacker obtains the \gls{TA} value from the RA response sent by the BS. These values are used to estimate the azimuth of the \gls{UE} and the distance from the \gls{BS} respectively, resulting in an estimate of the \gls{UE} location. Even though RA is primarily used for initial access, the attacker can also target already connected users by performing a PO spoofing attack to trigger the RA procedure, and estimate their location. We note that our attack does not require calibration or multiple devices for triangulation, and relies solely on measurements reported by the UEs. In Section~\ref{sec:results} we evaluate this attack by fingerprinting the beam locations in a real-world deployment, and measuring the localization error against our test device.

\subsection{Identified vulnerabilities in L2 procedures}
\label{sec:L2vulnerabilities}

As described in Section~\ref{sec:attackoverview}, an adversary is able to inject forged data headers of the unprotected \gls{PDCP}, \gls{MAC} and \gls{RLC} layers to a connected user.
\gls{PDCP} and \gls{RLC} headers contain control information for the transfer of upper layer data carried in their payload, e.g., how the information is compressed or concatenated. Thus, this header information does not control the UE behavior, rather how to process the encapsulated higher layer data. We recognize that spoofing these headers is possible, and has been exploited in previous work~\cite{CIoTSignalling}, however, in this work we focus on exploits on \gls{MAC} \glspl{CE} due to their capability of changing the state of the \gls{UE}, and influence \gls{BS} decision-making algorithms (e.g., scheduling). As each 3GPP release introduces new \glspl{CE}, we focus on the latest 4G and 5G releases~\cite{ts36321,ts38321}.
With a systematic analysis of each MAC CE and its implications on the functionality of the UE and \gls{BS}, we identify a set of vulnerabilities that we discuss in the following.

\subsubsection{Attacks on Carrier Aggregation} \gls{CA} has been widely deployed to provide high throughput gains by aggregating bandwidths at multiple cells operating at different center frequencies, denoted \glspl{SCell}. One of the main drawbacks of \gls{CA} is the considerable energy consumption at the \gls{UE}, which is monitoring the \gls{PDCCH} simultaneously in multiple frequencies, and transmitting periodic \gls{CSI} reports. \gls{CA} energy consumption measurements show that \glspl{UE} experience an average current increase of 79\% when activating one additional \gls{SCell}~\cite{CAcurrentusage}. To tackle this issue, \glspl{SCell} are activated and deactivated dynamically at the \gls{MAC} layer by the \gls{MAC} \gls{CE} \textit{SCell Activation/De-activation}, when high throughput delivery is required. This \gls{MAC} \gls{CE} consists of a bitmap which indicates which \glspl{SCell} are activated or de-activated at the UE. The deactivation process can be performed either by the \gls{MAC} \gls{CE}, or by expiration of a timer, \textit{sCellDeactivationTimer}, configured through \gls{RRC} with values varying from 20ms to 1.28s. If not specified, the de-activation timer is set to infinity~\cite{ts36331}.

As a result, a malicious SCell Activation \gls{MAC} \gls{CE} against a \gls{UE} stealthily forces it to use a considerably higher amount of energy. Unaware of this change, the \gls{BS} does not de-activate the active \glspl{SCell} through a \gls{DL} \gls{MAC} \gls{CE}. In Section~\ref{sec:results} we experimentally measure the potential of this attack in a real world setup by surveying the \textit{sCellDeactivationTimer} configuration for three major \glspl{MNO}.

Moreover, an attacker can de-activate \glspl{SCell} for a device that uses \gls{CA} without the knowledge of the \gls{BS}. This incurs a drastic throughput reduction at the physical layer, and even higher in the application layer. For instance, if the traffic is delivered over TCP, the data is multiplexed at the \gls{MAC} layer, and the throughput plummets due to retransmissions and out-of-sequence delivery.

\subsubsection{Tracking and localization using reference signals}\label{sec:srscsi} 
Wireless communications require frequent measurement and reporting of channel conditions for resource allocation optimization and to enable techniques such as beamforming. These channel measurements are performed over known reference signals in the downlink and the uplink channels. The reference signals can be transmitted periodically, semi-persistently, or aperiodically, and this is conveyed through protected RRC messages. Periodic signals are constantly transmitted, whereas semi-persistent reference signals are transmitted with a fixed period and can be turned on or off through a MAC CE (\textit{\gls{SP} SRS Activation/Deactivation}). Finally, aperiodic reference signals are triggered through a DCI and involve one single transmission.

\paragraph{\bf{\gls{SRS}}} The \gls{SRS} is  used for channel estimation across the full uplink band, and is the main candidate to carry UL massive MIMO pilots.
The importance of \gls{SRS} for an adversary is three-fold. First, as it is an \gls{UL} transmission that the \gls{BS} does not expect, it will interfere with other \gls{UL} transmissions scheduled by the \gls{BS}, either jamming user data or polluting other users' \gls{SRS}s, which will disrupt the measured channel state information. Secondly, an attacker can disable semi-persistent SRS transmissions by sending a deactivation MAC \gls{CE} to a UE, which will disrupt the channel estimation at the \gls{BS}, lowering the UE throughput considerably, particularly in beamforming scenarios. Third, \gls{SRS} is a predefined wideband Zadoff-Chu signal, which provides ideal cross-correlation properties. Although the purpose of this signal is not for positioning, an attacker can use it to localize a specific user with high accuracy, by measuring the time difference of arrival~\cite{toaSRS}. 

\paragraph{\bf{Leakages in \gls{CSI-RS}}} The procedure for downlink channel measuring is slightly different than SRS. The \gls{CSI-RS} is transmitted periodically in the downlink by the base station, and the UE reports the measured \gls{CSI} back to the BS for scheduling purposes, as described in Section~\ref{sec:pucch}. The CSI report can be configured to be periodic, semi-persistent or aperiodic, similarly to the \gls{SRS}. An attacker can use the \gls{MAC} \gls{CE} \textit{\gls{CSI} reporting on PUCCH Activation/Deactivation}, to instruct the \gls{UE} to transmit - or stop transmitting - semi-persistent \gls{CSI} reports to the \gls{BS}. Although this can create UL collisions between users, the absence of \gls{CSI} reports is detectable by the \gls{BS}.
However, we find that the \gls{CSI} report contains crucial information that can be leveraged by a passive attacker to track users' movement in a cell. 
Specifically, in beamforming scenarios, the CSI includes a pair of values, $\{B_{idx},RSRP\}$, where $B_{idx}$ is the beam index identifier of the measured strongest beam, and $RSRP$ is the measured signal strength for given beam. This enables swift beam management, as the \gls{BS} will use this information to transmit the downlink information to a UE using the strongest beam. The CSI report is sent in the clear, carried by the PUCCH, and contains the RNTI of the UE. 

This information can be used by an attacker to track users' location in three steps: First, an attacker fingerprints the static cell beam configuration, i.e., measures the physical area covered by each beam index. This step is shared with our beamforming localization attack described in Section~\ref{sec:prach}. Second, the attacker decodes the PUCCH, retrieving the pair $\{B_{idx},RSRP\}$ from the CSI reports and the RNTI of the UE.
Finally, the attacker estimates the GPS coordinates that describe the UE path from the beams reported by the UE, using our path inference algorithm, described in \texttt{BeamToPath} (Algorithm~\ref{alg:csipath}).
Our algorithm compensates for various factors that affect wireless propagation, such as reflections, blockages and non-line-of-sight, using a three-step filtering process. First, Power Filtering identifies outliers in RSRP values, e.g., high variations in a very short time, due to blockages. Second, the Smoothening filter measures the consecutive reported occurrences of each beam, and discards very low frequency beams. This can be due to sporadic reflections, but also due to transitions between two contiguous beam, which results in the path quickly flickering between two beams. Lastly, the algorithm discards impossible beam transitions, e.g., moving from beam A to beam B where beam A and B are not contiguous. Finally, the algorithm obtains the coordinates of the path from the coordinates of the center of each reported beam, and interpolates the path using linear interpolation. In Section~\ref{sec:results} we present our attack evaluation in an urban scenario by tracking the movement of our test device connected to an operator network. 

\begin{beamToPath}
\begin{algorithmic}
    \State \textbf{Input:} \{$\{({B_{idx,i},RSRP_i})\}_{1..N}, RSRP_{base}$\}
    \State \textbf{Output:} Set of latitude/longitude coordinates (path)
    \State $count \gets 0~\text{for all idx}$;
    \For{$n = 0,\ldots,N$} 
        \If{$|RSRP_n - RSRP_{base}| > P_{thres}$}
        \State continue;
        \EndIf
        \If{$B_{idx,n} == B_{idx, n+1}$}
            \State $count++$;
            \Else
            \If{($count > C_{thres}$) $\&\&$ \\ areContiguous($B_{idx}$, ${Path}[-1]$)} \State addToCoordinates(GPSCoords($B_{idx}$));
            \State $count \gets 0$;
        \EndIf
        \EndIf            
    \EndFor
    \State return linearInterpolation(Coordinates);
\end{algorithmic}
\caption{Infer path from CSI reports}\label{alg:csipath}
\end{beamToPath}

\subsubsection{Beam Failure Recovery attack} 
Beam management needs fast, dynamic reconfiguration to adapt to sudden changes in the wireless medium, such as signal blockage. 
Generally the base station initiates beam management mechanisms, such as beam swapping, however, when a UE detects beam failure, it measures the strongest beam and initiates RA including \gls{BFR} MAC CE in \texttt{Msg3}, indicating that the user is changing to a different beam. This procedure can be spoofed by an attacker, creating misalignment between the beams at UE and BS, as the legitimate UE is unaware of the beam switch.

\subsubsection{Other MAC CE vulnerabilities}

Previous work has described injection attacks on a small set of MAC CEs, which we do not include in our evaluation, as we focus on new vulnerabilities. Prior work reported the Buffer Status Report (BSR), Discontinuous Reception (DRX), and Power Headroom (PHR), theoretically~\cite{bsrinjection2007}, and in a C-IoT testbed~\cite{CIoTSignalling}. Other MAC CEs that lead to service degradation, with less overall impact are Recommended BitRate, which can degrade the QoS of users by spoofing a low recommended bitrate request from the UE, or Timing Advance (\gls{TA}) MAC CE, which can be injected to a UE to de-synchronize the uplink transmissions. More importantly, TA will be used as one of the inputs of our localization attack described in Section~\ref{sec:prach}.

\begin{figure*}[ht]
\begin{center}
\includegraphics[scale=0.2]{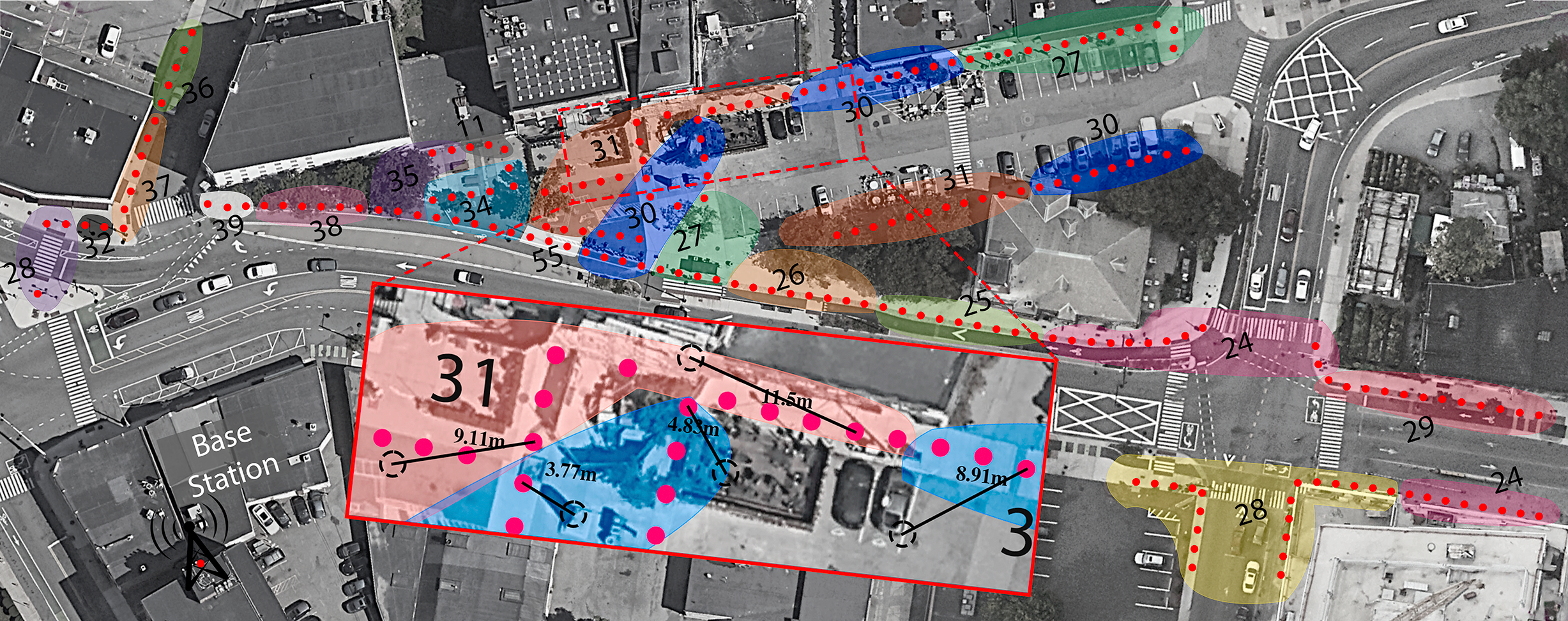}
\end{center}
\caption{Fingerprinting of static beams within a cell. Red points denote the GPS locations of every measurement and the BS. Colored overlays denote the fingerprinted beam locations. The magnified area shows five localization examples with annotations of the distance between the actual and the estimated locations.}
\label{fig:beamsnionsquare}
\end{figure*}

\subsection{Technical challenges and practical considerations}
\label{sec:technicalchallenges}

In the following, we discuss the challenges of successfully spoofing control information to a connected UE in a real-world scenario, and the ways to overcome them.

\paragraph{\bf{Over the air injection}} 
Since 2G systems, spoofing data to a cellular user was achieved using a rogue \gls{BS}. However, this approach has three main drawbacks: 1) it requires high power, 2) the rogue \gls{BS} does not have the operator keys and the mutual authentication procedure with the \gls{UE} fails, and 3) it is not a stealthy approach, as defenses are currently in place in the standard~\cite{ts33501}.  
Recently, researchers have demonstrated a stealthy, power-efficient alternative, spoofing messages to a \gls{UE} by overshadowing legitimate transmissions, using only 3dB higher power than the BS~\cite{hidingplainsignal}. This has been shown to be possible both in the \gls{DL}~\cite{hidingplainsignal, LTrack, sigunder} and in the \gls{UL}~\cite{adaptOver}. Practical challenges such as time and frequency synchronization are overcome leveraging the length of the cyclic prefix, the loose time synchronization requirement provided by OFDM, and the overshadowing of reference signals. 

\paragraph{\bf{\gls{RRC} Configuration at the \gls{UE}}} One of the limitations of spoofing \gls{MAC} \glspl{CE} is that the attacker is oblivious to the static \gls{RRC} configuration of a user, as it is conveyed encrypted. Hence, the attacker does not know which MAC CEs the phone processes dynamically. A passive approach can be used by sniffing the headers to reconstruct the current state of the UE, with open-source tools such as~\cite{yongdaesniffer}. Alternatively, the UE behavior can be monitored
after being probed with various MAC CEs.

\paragraph{\bf{Obtaining radio user identifiers}} Our attacks are targeted and as such assume that the attacker can obtain the \gls{RNTI} of the victim. Decoding the active \glspl{RNTI} in the \gls{PDCCH} has been shown to be possible since 4G~\cite{lteeye,OWL,FALCON}, and recently in 5G~\cite{5gsniffer}. Although the \gls{RNTI} is a temporary identifier, which does not link to a permanent identity, our attacks can be made targeted, as researchers have shown that it is possible to map identities, such as a phone number, to \glspl{RNTI} in 4G~\cite{breakinglayertwo,ltelocationtracking} and 5G~\cite{5gsniffer}.

\section{Experimental Evaluation}
\label{sec:evaluation}

In this section we present our experimental results that evaluate attacks described in Section~\ref{sec:attack} and summarized in Table~\ref{table:summarizedattacks}. Note that active attacks are evaluated in our own isolated anechoic chamber testbed. We were not able to experimentally validate some active attacks for ethical reasons and because the target functionalities such as \gls{BWP} Switching or Beam Failure Recovery are not fully supported by open-source implementation or COTS UEs. Additionally, we carry out a passive-measurements survey of MNO network configurations across three different countries to validate if operators use configurations that make users more susceptible to given attacks.

\subsection{Evaluation setup}
\label{sec:evaluationsetup}
Our evaluation setup consists of three distinctive scenarios, passive UE localization attacks, active injection attacks and network configuration surveying.
For our passive localization attacks, we attach our COTS UE (Google Pixel 5) to the operator network using multiple beams in an urban location and log using QXDM~\cite{qxdm} relevant information that is being sent unprotected. 
For active injection attacks, we perform our measurements using a modified version of srsRAN~\cite{srsRAN}, where we integrate the attacker in the base station code.
For UE devices we use a COTS UE, Pixel, 5 with custom SIM cards, as well as the srsRAN UE implementation. For SDR we use the Ettus USRP B210 and X310.
To survey the current configuration of \glspl{MNO}, we use our COTS UE with SIM cards of the different \glspl{MNO}, and obtain the network information during UE operation.

\textit{\textbf{Ethical Considerations}.}
Our evaluation is performed following strict ethical guidelines. Our passive and network surveying attacks are performed by obtaining information from our device with a valid SIM card. We note that connecting our COTS UE to the network and logging the RRC and PHY messages is normal operation and does not disrupt the networks. 
For our active attacks that involve over the air transmissions, we perform them on our own isolated testbed and make sure there is no interference
with regular systems. All the RF transmissions from our srsRAN base station and the UEs are performed inside an anechoic chamber, that completely isolates RF emissions from the outside world. We verify this with spectrum measurements.

\subsection{Results}
\label{sec:results} 

We experimentally evaluate our beamforming-based localization and tracking attacks, DCI spoofing, including \gls{HARQ} injection, barring connected and not connected UEs from accessing the network in RA attacks, and attacks related to Carrier Aggregation.

\subsubsection{\bf{Passive user localization leveraging beamforming leakages}} 
In this subsection we evaluate our attacks on beamforming that lead to passive user localization (Section~\ref{sec:prach}) and user movement tracking (Section~\ref{sec:srscsi}).
We find that even small amounts of data, such as fetching email or sending one stealthy Signal or Telegram message~\cite{5gsniffer} triggers the use of mmWave/beamforming, making users highly susceptible to the attack.

\paragraph{\bf{Fingerprinting beam locations}} To perform our attacks, we initially fingerprint the locations of the beams used in an urban environment by a 5G operator. The 5G BS (situated on top of a building, at 23 meters height) operates in the 28 GHz band, with a subcarrier spacing of 120~KHz, and uses 48 beams to cover the area within an angle of approximately 120 degrees. We connect a COTS UE to the base station, and log the beam index that our UE reports as well as the TA value reported by the BS. We repeat this process while walking within the coverage of the base station, and we map the areas served by different beams. Our fingerprinting comprises the beam areas, the distribution of TA values within each beam area, and one line for each beam area that starts from the BS location and splits the area in half. The locations of the measured beams are depicted in Figure~\ref{fig:beamsnionsquare}. We note how the shape of the areas depends on the line of sight from the BS, and they span a width of approximately 20 meters. Additionally, in the absence of line of sight, beamforming coverage becomes unstable and is either lost, or the reflections of other beams are dominant and serve the UE. This is the case with the beams marked as 30, 31, and 24 in Figure~\ref{fig:beamsnionsquare}. However, in the area right in front of the base station there is no beamforming coverage due to obstruction by buildings.
Finally, we verified that this beam configuration is static, i.e., remains the same throughout several experiments conducted months apart. We use the output of this fingerprinting phase for both our localization and movement tracking attacks described below.

\begin{figure}
\centering
\includegraphics[scale=0.8]{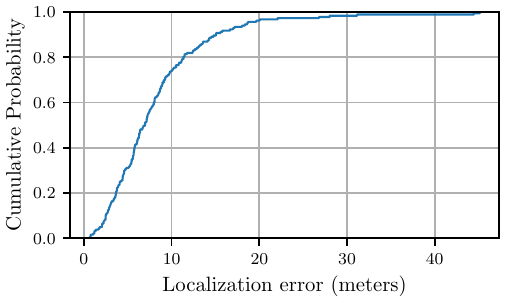}
\caption{Empirical cumulative distribution function of the localization error across all points in our dataset.}
\label{fig:ssbraecdf}
\end{figure}

\paragraph{\bf{SSB-RA localization}} In order to evaluate our localization attack, we position our COTS UE (target of the attack) in the coverage of the BS and we connect to the cell. We log the beam index used for RA, the TA value sent by the BS in the RA response, and the exact GPS location. We take measurements every 2 meters across the beamforming coverage of the BS, and obtain a dataset with a total of 183 points. For each point, we compute the estimated user location based on our fingerprinting, using the beam index and the reported TA.
First, the reported beam index indicates the azimuth angle of our location estimate, i.e., the line that starts from the BS and splits the beam area in half. When multiple areas are present (e.g., beam 30 in Figure~\ref{fig:beamsnionsquare}), we choose the area that better matches the reported TA value based on our fingerprinted distribution of TA values within the beam areas. 
Second, we translate the reported TA value to an estimate of the UE distance from the BS using the formula $T_{TA}=f(TA)$ from Section 4.3.1 in~\cite{ts38211}. In the case of the cell configuration in our experiment, one increment corresponds to a distance increase of $9.77$ meters. Due to the discretized design of TA values, instead of a single distance estimate we obtain the range of distances $[f(TA-1), f(TA)]$, which defines an annulus (or ring shape) around the BS. We compute the line segment that is the intersection of the beam line and the TA annulus, and output the midpoint of this line segment as our location estimate.
Finally, we compute the localization error as the distance between our estimate and the ground truth location, and show the ECDF of the error in Figure~\ref{fig:ssbraecdf}. Our results show that an attacker can localize a user with an error less than 10 meters 74.06\% of the time, and less than 20 meters 96.01\% of the time, regardless of their location in the cell, by passive sniffing the RA exchange.

\begin{figure}
\centering
\includegraphics[scale=0.5]{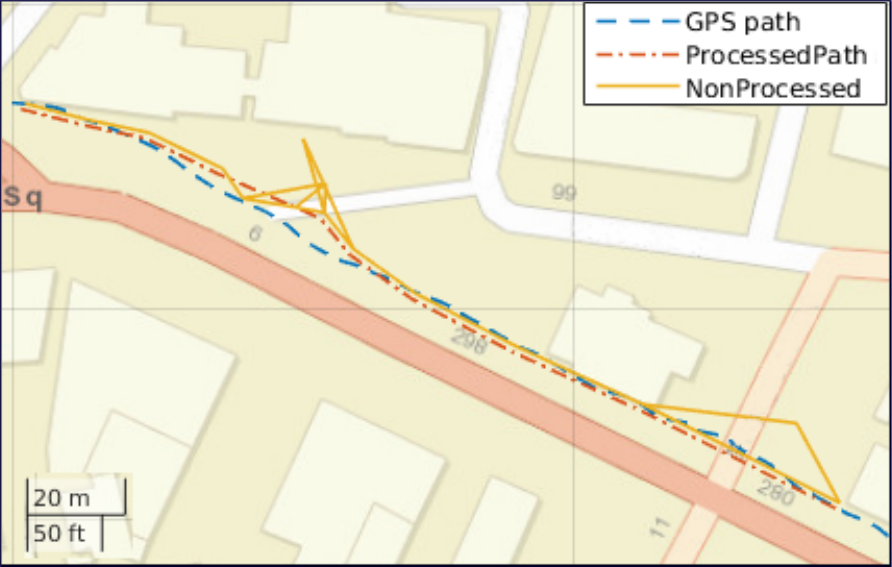}
\caption{Comparison of the GPS path and the estimated path using CSI Reports, with and without our filtering algorithm.}
\label{fig:filteringpaths}
\end{figure}

\paragraph{\bf{User movement tracking with CSI reports}} To validate and evaluate our tracking attack by leveraging leakages in CSI reports, we connect our COTS UE (target of the attack) to the operator network, and walk in different path patterns, including changes of direction, covering the area that the mmWave cell serves. Our total dataset comprises more than 2 km in path lengths ranging from 20 meters to 150 meters, with a total of 30 paths, that cover all the pedestrian area. We log both the COTS UE GPS position and the CSI reports sent by the UE from the chipset. We use the sequence of CSI reports as input to the algorithm described in Section~\ref{sec:srscsi}, which outputs the estimated path that the UE traversed. Figure~\ref{fig:filteringpaths} depicts one of our paths taken. The figure includes the GPS (groundtruth) path, and our estimated path using CSI reports with and without being processed by our filtering algorithm, as described in~\ref{sec:srscsi}. We find that the non-filtered path presents high fluctuations, commonly close to areas where the density of beams increases. More importantly, we find that the path described by the non-filtered path quickly flickers between points when the user is transitioning from one beam to the next. This makes the distance covered by the non-filtered estimated path considerable larger than the distance covered by the GPS path, and does not reflect the real path taken by the user. In comparison, our filtered path smoothly follows one single path, and the total distance of the described path is within 15 meters of the measured GPS path. By using our algorithm, the attacker is able to successfully track the movements of a specific user within a cell.  In order to evaluate if the path is recovered correctly, we compute the maximum distance deviation between the estimated path and the ground truth GPS path, for all our recorded paths. Our results show that we are able to recover the correct path with a maximum deviation below 14 meters in 90.32\% of the paths, and the average error across paths is 5.34 meters. Moreover, we measure the CSI report configurations in cells supporting beamforming across six MNOs spanning three different countries. We find that, to support beamforming, all operators use periodic CSI configurations, every 20 to 40 ms, which provide fine granularity.

\begin{figure}
\centering
\includegraphics[scale=0.8]{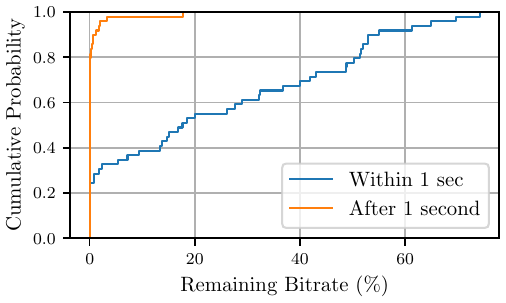}
\caption{Empirical cumulative distribution function of Bitrate DoS with DCI Spoofing.}
\label{fig:dci_spoof_ecdf}
\end{figure}

\subsubsection{\bf{DCI Spoofing}}
In the following, we evaluate our DCI spoofing attacks, that lead to UE collisions, service degradation, and denial of service. We craft DCI messages and inject them to active UEs, which cannot detect the injections since the messages are not integrity protected.

\paragraph{\bf{Tricking legitimate UEs to jam other UEs}}
To validate the theoretical attack, we initially inject DCIs with UL grants of various sizes to active UEs that do not have any data to transmit, and are not requesting any resources. We find that the UEs use all the resources allocated to them, and fill the allocated space with padding. We setup the attack evaluation by having two COTS UEs connected to our 5G testbed. The first UE generates Iperf3 traffic on the uplink at maximum bandwidth, and will be the target of our attack. The second UE, referred to as Induced Jammer UE (IJ-UE), will be tricked to jam. We inject a crafted DCI message to the IJ-UE and set the UL Grant allocation to include all the physical resource blocks of the \gls{PUSCH} during every time slot. Additionally, we set the \gls{TPC} parameter to the highest value, which instructs the UE to increase the transmit power to the maximum. We monitor the throughput reported by Iperf throughout the duration of the attack over 1-second intervals and we measure the average throughput: before the attack, during the first second after the attack, and during the remaining duration of the attack after the first second.
Figure~\ref{fig:dci_spoof_ecdf} shows the ECDF of the affected throughput as a percentage of the original throughput of the UE before the attack. We notice that 77\% of the time the target UE achieves less than 50\% of its original throughput within one second of the attack and less than 0.1\% throughput for the remainder of the attack. Therefore, an adversary is able to trick UEs against each other, blocking all communications, at the cost of spoofing very small DCI messages. 

\paragraph{\bf{\gls{HARQ}-Attack}} We validate the attack described in Figure~\ref{fig:harqattack} against a COTS UE. To do so, we inject a DCI with an out-of-sequence DAI, and we analyze the reported ACK bitmap. We find that by injecting a DCI with a DAI counter $n$ above the current counter value, the UE reports an ACK report with $n$ more bits than the base station expects, accounting for the $n$ missed transmissions. For instance, by injecting a DAI with value 2, the UE reports 3 bits, with two NACKs corresponding to DAIs 0 and 1. To evaluate our attack, we connect our COTS UE to our 5G srsRAN BS, and we spoof DCIs with a DAI value above the current DAI. We observe that the base station is unable to match the received ACK bitmap to the correct transmissions, and reaches HARQ failure. Consecutive missed ACKs lead to a radio link failure after two seconds, and breaks the connectivity between the COTS UE and the base station. Due to ethical reasons, we do not evaluate the attack on an operator network, but we highlight that bitmap mismatches are not handled by the 3GPP standard.

\subsubsection{\bf{Blocking cell access through RA Attacks}} As described in Section~\ref{sec:prach}, RA parameters are broadcast and optimized for traffic management and flexibility at the \gls{BS} side. First, we survey MNOs configurations, and in Table~\ref{table:RRC_params}, we present the measured RA configuration of six major MNOs across three countries (totalling over 500 million subscribers). We find that the maximum number of \gls{RA} attempts is set to only $5$ or $10$, and the \gls{RAR} window size is set to the maximum allowed, i.e., \texttt{sf10}. This reduces the load on the random access channel, and loosens the tight requirement on response times for the BS. 

\begin{figure}
    \centering    \includegraphics[width=0.7\columnwidth]{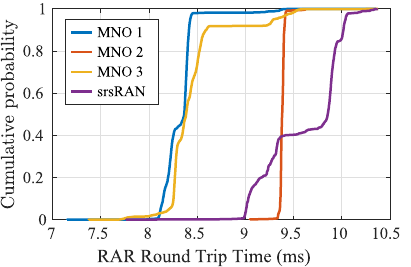}
    \caption{Empirical cumulative distribution function of the measured RAR RTT time for three major \glspl{MNO}.}
    \label{fig:ecdfRAR}
\end{figure}

To verify that a modified window size successfully leads to RA failure, we compute the empirical distribution of the RAR \gls{RTT} per \gls{MNO} in Country A. We measure the time duration between the transmission of
the \texttt{RA Preamble (Msg1)} by the client, and the receipt of \texttt{RAR (Msg2)} by the client, using QXDM~\cite{qxdm} to extract the logs from the Qualcomm chipset. We repeat this experiment over three days and collect data for more than 1000 \glspl{RA} per \gls{MNO}. Figure~\ref{fig:ecdfRAR} shows the \gls{ECDF} of the RAR messages for all three \glspl{MNO}, and our srsRAN testbed. Our results show that  RAR RTT times have a fixed value with very low variance, and are consistently higher than 7ms (which corresponds to 7 subframes), indicating that if the SIB2 injection forces a window size of less than 5 subframes, \glspl{RA} will systematically fail. To confirm this, we set the \texttt{ra-ResponseWindowSize} and the
\texttt{preambleTransMax} parameters in the SIB of our isolated srsRAN  BS to \texttt{sf2} and 200 respectively, without the \gls{BS} being aware, and we attempt connecting our COTS UE. The \gls{RA} during initial access fails 200 times, and repeats periodically every few seconds with the same result.

\begin{figure}
    \centering    \includegraphics[width=0.7\columnwidth]{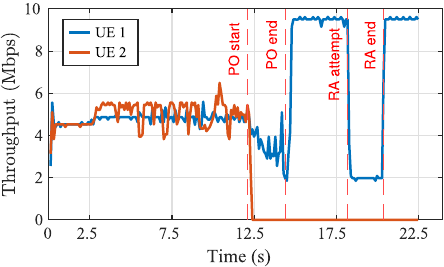}
    \caption{Measured Iperf3 \gls{TP} from two UEs during the PDCCH Order (PO) amplified attack, starting at instant t=12 for UE 2.}
    \label{fig:iPerfPDCCHOrder}
\end{figure}

However, this only bars connection to new users, but does not impact connected users. In order to target connected users, we leverage the \gls{PO} command, and  measure the impact on the targeted victim, and on non-targeted UEs served by the same \gls{BS}. We setup our \gls{BS} and connect two UEs. We generate TCP traffic and monitor the throughput using iPerf3 on both devices, during which we inject a \gls{PO} to one of the devices. The measured iPerf \gls{TP} for both devices is depicted in Figure~\ref{fig:iPerfPDCCHOrder}. Our results show that, at first, both UEs generate traffic and share the spectrum, achieving around 5~Mbps  each. However, when the \gls{PO} is injected to UE2, its throughput becomes 0, as it is performing multiple (failing) RAs.
The 200 RAs last 2.1~seconds, leading to Radio Link Failure for UE2, which disconnects from the network and switches to idle. Therefore, UE1 uses all resources in the channel and doubles its \gls{TP} to almost 10~Mbps. However, when UE2 tries to re-connect to the cell, after 3 seconds, the RA flood affects the already connected UE, whose throughput drops by 79\%.
\ignore{This can be a result of both the computational stress on the \gls{BS}, as well as the fact that} This is likely due to \gls{RAR} responses using uplink and downlink resources, reducing the total available resources for other UEs.

\subsubsection{\bf{Carrier aggregation attacks}} 
\begin{figure}
    \centering    \includegraphics[scale=0.8]{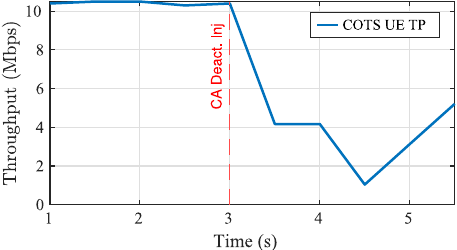}
    \caption{Measured Iperf3 \gls{TP} from a UE during a SCell Deactivation MAC CE injection at t=3.}
    \label{fig:iPerfSCellDeact}
\end{figure}

We configure our LTE BS to operate over two 5 MHz carriers in LTE bands 3 and 7. We connect our COTS UE with enabled \gls{CA} over two carriers. We generate downlink TCP traffic using iPerf3 and monitor the throughput between the \gls{UE} and \gls{BS}. We then spoof a SCell Deactivation DL MAC CE to the COTS UE at time instant 3 seconds, instructing the UE to deactivate the SCell. The measured iPerf TP is depicted in Figure~\ref{fig:iPerfSCellDeact}. The UE maintains a throughput of 10.5 Mbps using \gls{CA} for the initial 3 seconds, but this value rapidly decreases to as low as 1.04 Mbps, at the time of the attack. The observed TCP retransmissions and the throughput drop to below 50\%, validate that the UE no longer monitors the secondary cell that the BS is still using to transmit iPerf traffic.

In order to assess the efficiency of our attacks in real-world scenarios, without targeting deployed networks, we analyze the current configuration of three \glspl{MNO} serving over 100 million subscribers each.
We use QXDM to extract the configuration parameters of interest for \gls{RA} from SIB and \gls{CA} from the RRC by connecting a phone with the \gls{MNO}'s SIM card, and include our findings in Table~\ref{table:RRC_params}.

\begin{table}[]
\setlength{\belowrulesep}{0.5ex}
\setlength{\aboverulesep}{0.5ex}
\begin{center}
\begin{tabular}{@{}llllll@{}}
\toprule
& MNO      & SCellDeact. & Max CA & Max RA &  RARWind\\ \midrule
& \gls{MNO} 1 & Infinite                     & 3CA & n10                       & sf10       \\     
Country A & \gls{MNO} 2  & Infinite    & 4CA     & n5                        & sf10     \\         
& \gls{MNO} 3    & Infinite & 5CA & n10  & sf10         \\ \bottomrule  
\multirow{2}*{Country B} & \gls{MNO} 1 & Infinite                     & 3CA & n10                       & sf10       \\     
 & \gls{MNO} 2  & Infinite    & 3CA     &  n10                       & sf10     \\ \bottomrule       
{Country C} & \gls{MNO} 1 & Infinite                     & 3CA & n10                       & sf10       \\   \bottomrule

\end{tabular}
\caption{Observed values of LTE RRC parameters related to CA and RA from six major \glspl{MNO} across three countries.}
\label{table:RRC_params}
\end{center}
\end{table}

One factor that decreases the effect of MAC CE injection attacks is the presence of timeouts for various
UE states, such as the \texttt{sCellDeactivationTimer}. The goal of our \gls{SCell} activation attack is for the UE to maintain the \glspl{SCell} (that the attacker enforces) active for an extended period of time, to avoid frequent injections. Table~\ref{table:RRC_params} shows that, in practice, all three \glspl{MNO} choose to omit the timer value in the RRC configuration, leaving it to the default (i.e., infinity). This simplifies the control process for the BS because
it has full control of \gls{SCell} activation/deactivation through MAC CEs, without accounting for UE timers, but makes
\glspl{UE} more vulnerable and leads to considerable battery draining.
We experimentally verify that the UE successfully activates a configured \gls{SCell} upon a SCell Activation DL MAC CE injection by the attacker.
Moreover, we find that all tested MNOs support between 3 and 5 aggregated carriers, and according to previous work, each activated carrier contributes an average current increase of 79\%~\cite{CAcurrentusage}.

\section{Mitigations}
\label{sec:defenses} 
Lack of protections of low-layer control procedures leads to a range of privacy and security attacks. This calls for a careful redesign of these mechanisms. In the following we discuss potential defenses and mitigations.

\BfPara{Techniques against spoofed signalling} Spoofing and sniffing of low layer information can be mitigated by extending encryption to the lower layers. To do so, the UE and base station agree on a new key for L1 and L2, $K_{PHY}$,  derived after authentication, with a process similar to keys derived for  RRC layer~\cite[p.~48]{ts33501}. In order to minimize changes to the standard, we do not add additional processing blocks for encryption, but propose to modify the existing scrambling block, present in all physical channels, e.g., PDCCH/PUCCH, PDSCH/PUSCH. This way, instead of scrambling the data with a sequence derived from known parameters, the BS encrypts the physical channel information with a cryptographic sequence generated using as input the derived key, $K_{PHY}$, the system frame number, $\texttt{SFN}$, subframe, $\texttt{sf}$, and starting index of resources blocks used for the transmission, $\texttt{RB}$. Practically, this is equivalent to using encryption (e.g., AES) with Counter Mode. It has the additional advantage of allowing precomputation of the scrambling/encryption sequences, thus minimizing delay.

A less intrusive approach consists of using lightweight message authentication mechanisms. For instance, digital signatures can be used to guarantee the integrity of SIBs, using the Non-Critical Extensions field (NCE). This would for example protect against attacks exploiting \texttt{ra-ResponseWindowSize}, and \texttt{preambleTransMax}.
Integrity information of MAC CEs and DCI messages, can be incorporated in upper (protected) layers at the expense of increased overhead and latency. The integrity can even be verified asynchronously to allow for faster reconfiguration, in particular when the effect of the attack can be undone (e.g., activation/deactivation of secondary cells). An additional defense consists of attacks detection either by the BS or the UE. Overshadowing can be detected by the BS with successive interference cancellation~\cite{sigunder}, or by monitoring broadcast transmissions like SIBs to detect rogue messages, or by requesting the UE to report a decoded SIB upon connection. Additionally, the UE can send acknowledgments for the received MAC CEs or DCI messages like PDCCH Order or BWP Switching.

\BfPara{Passive user localization} Our localization and tracking attacks are possible due to three design choices, i) the beams remain static for long periods of time, ii) the CSI reports are unprotected and indicate the strongest measured beam index, and iii) there is a one-to-one mapping of Random access occasions and strongest measured beam. As operators carefully arrange the beams in the best configuration for a specific area, re-arranging dynamically would require considerable effort, and it would not completely prevent fingerprinting. For this reason, our proposed mitigations focus on the two other points. For point ii), we note that CSI reports are configured after RRC connection, and as such, it is possible to leverage the existing security context. For instance, it is possible to convey a key through an encrypted RRC message, or use the key $K_{PHY}$ that we describe earlier in this section. By using this shared secret, UE and gNB can encrypt the full CSI report. In order to minimize overhead, yet guarantee IND-CPA security, we propose to use time, $\texttt{SFN}$, and $\texttt{sf}$ as the Initialization Vector for  encryption.
\ignore{or individually the beam indices, i.e. derive a mapping between reported beam indexes and real beams, that an attacker can not resolve. In this way, an attacker would see a sequence of strongest beam indices, but it would not be able to retrieve the mapping to physical locations. In a similar way, our RA localization attack can be mitigated by establishing an initial connection through a PCell, and informing through a protected RRC connection, an encrypted mapping (per UE) between RA occasions and beam indices, which would deem a fingerprinting attack ineffective.}

Finally, the 3GPP standard can be updated to enforce fallback configurations and default parameter values, to support
switching to a less vulnerable mode when an attack is detected. For instance, from our surveying of information, configuration parameters such as Infinite timer for secondary cell de-activation make the UEs more vulnerable to data signalling spoofing attacks, similarly safe default values for \texttt{ra-ResponseWindowSize} and \texttt{preambleTransMax} make the \gls{RA} procedure robust against spoofed SIBs.

\section{Related Work}
\label{sec:relatedwork} 
\BfPara{Spoofing wireless signals}
Impersonation and data spoofing has proved to be a major concern in cellular networks. The lack of integrity protection at different layers~\cite{rootofallevil} enables the existence of rogue base stations and sophisticated MiTM attacks that lead to impersonation and data spoofing attacks~\cite{imp4gt, breakinglayertwo, lteinspector, practicalattacks, expdevicecapabilities}. In recent years, a new stealthy approach, spoofing attacks by overshadowing legitimate signals was introduced in~\cite{hidingplainsignal}. The authors show that it is possible to spoof unprotected broadcast \gls{DL} RRC messages such as SIB, by transmitting at a slightly higher power than the BS. The authors showcase DoS and cascading attacks leveraging such technique. Since then, additional work expanded this concept, including uplink overshadowing~\cite{adaptOver}, and low-power stealthy overshadowing of broadcast signals~\cite{sigunder}. 

\BfPara{Exploiting low layer mechanisms} The lack of ciphering and integrity protection at low layers of the protocol enables spoofing and sniffing of user data. However, research on active attacks on low layers of cellular networks is scarce. Authors in~\cite{bsrinjection2007} identified the threat of exploiting Buffer Status Report (BSR) MAC CEs and leveraging the long Discontinuous Reception (DRX) state for message insertion, but do not provide practical evaluation. Authors in~\cite{CIoTSignalling} show a practical evaluation of \gls{MAC} \gls{CE} forging attacks on C-IoT devices, which leverage details specific to the C-IoT physical layer and \gls{MAC} \glspl{CE}, but state that the attacks are not applicable to broadband cellular systems. Some of the shown attacks disrupt the UE connectivity by targeting the RLC layer ACKs, the DRX procedure and BSR. Mitigation techniques in the literature include attack detection by verifying the data signaling correctness~\cite{cellDAM}, or expanding the integrity and ciphering to MAC and RLC layers~\cite{CIoTSignalling}. Our exhaustive analysis evaluates all low layer procedures and physical channels, including new 5G techniques and procedures and tens of recently introduced \gls{MAC} \gls{CE}s, uncovering several new attacks not described in previous literature, in particular those combining multiple mechanisms. 

\BfPara{Victim localization and tracking} Practical location attacks have been shown in the literature for all network technologies since GSM networks~\cite{2glocation,torpedo, 5gsniffer,smslocation}, for instance by determining the presence of a user within a cell based on broadcast messages. More accurate localization attacks have been shown recently in 4G networks, by leveraging the 
measured \gls{TA} MAC CE by the BS, and measuring the TA from another calibrated hardware, triangulating the UE position. Additionally, the authors in ~\cite{scelltracking} leverage the \gls{SCell} Activation/Deactivation \gls{MAC} \gls{CE} to fingerprint users' paths based on the \glspl{SCell} they are connected to, and evaluate the attack in an emulated WiFi scenario. To the best of our knowledge, our passive localization attacks are the first to exploit 5G-specific beamforming design vulnerabilities, identify combined attacks with other mechanisms (e.g., \gls{PO}), and is the first evaluation in real 5G networks.


\end{document}